\newcommand{\NPA}[3]{Nucl.\ Phys.\ A\ {\bf #1},\ #2 (#3)}

\newcommand{\PLB}[3]{Phys.\ Lett.\ B\ {\bf #1},\ #2 (#3)}

\newcommand{\PRL}[3]{Phys.\ Rev.\ Lett.\ {\bf #1},\ #2 (#3)}

\newcommand{\PRC}[3]{Phys.\ Rev.\ C\ {\bf #1},\ #2 (#3)}
\newcommand{\PRD}[3]{Phys.\ Rev.\ D\ {\bf #1},\ #2 (#3)}

\newcommand{\EPJC}[3]{Eur.\ Phys.\ J.\ C\ {\bf #1},\ #2 (#3)}

\newcommand\s{\sigma}

\newcommand{\diracslash}[1]{#1\llap{/\kern2pt}}

\newcommand{\be}{\begin{equation}}
\newcommand{\ee}{\end{equation}}
\newcommand{\bea}{\begin{eqnarray}}
\newcommand{\eea}{\end{eqnarray}}
\newcommand{\ba}[1]{\begin{array}{#1}}
\newcommand{\ea}{\end{array}}

\documentclass[prd,aps,floats,nofootinbib,tightenlines,showpacs]{revtex4-1}
\usepackage{epsfig,graphicx,pstricks}
\usepackage{psfrag}
\usepackage{color}
\usepackage{amsmath}
\usepackage{amsfonts}
\usepackage{amssymb}
\usepackage{textcomp}
\usepackage{multirow}

\begin{document}

\title {Bulk and shear viscosities of hot and dense hadron gas}
\author{Guru Prakash Kadam }
\email{guruprasad@prl.res.in}
\affiliation{Theory Division, Physical Research Laboratory,
Navrangpura, Ahmedabad 380 009, India}
\author{Hiranmaya Mishra}
\email{hm@prl.res.in}
\affiliation{Theory Division, Physical Research Laboratory,
Navrangpura, Ahmedabad 380 009, India}

\date{\today} 

\def\be{\begin{equation}}
\def\ee{\end{equation}}
\def\bearr{\begin{eqnarray}}
\def\eearr{\end{eqnarray}}
\def\zbf#1{{\bf {#1}}}
\def\bfm#1{\mbox{\boldmath $#1$}}
\def\hf{\frac{1}{2}}
\def\sl{\hspace{-0.15cm}/}
\def\omit#1{_{\!\rlap{$\scriptscriptstyle \backslash$}
{\scriptscriptstyle #1}}}
\def\vec#1{\mathchoice
        {\mbox{\boldmath $#1$}}
        {\mbox{\boldmath $#1$}}
        {\mbox{\boldmath $\scriptstyle #1$}}
        {\mbox{\boldmath $\scriptscriptstyle #1$}}
}

\begin{abstract}
We estimate bulk and shear viscosity at finite temperature and
baryon densities of hadronic matter within  hadron resonance gas model.
For bulk viscosity we use low energy theorems of QCD for the energy
momentum tensor correlators. For shear viscosity coefficient, we estimate
the same using molecular kinetic theory to relate the shear viscosity
coefficient to average momentum of the hadrons in the hot and dense
hadron gas. The bulk viscosity to entropy ratio increases with chemical
potential and is related to the reduction of velocity of sound at nonzero
chemical potential. The shear viscosity to entropy ratio on the other hand, 
shows a nontrivial behavior with the ratio  decreasing with 
chemical potential for small temperatures but increasing with chemical 
potential at high temperatures and is related to decrease of entropy density
 with chemical potential at high temperature due to finite 
volume of the hadrons.
\end{abstract}

\maketitle

\section{Introduction}
Recently, the transport properties of hot and dense matter has attracted 
lot of attention in the context of relativistic heavy ion 
collisions\cite{heinzrev}
as well as cosmology\cite{berera}. Such properties enter in the hydrodynamical
evolution and therefore essential for studying the near equilibrium 
evolution of a thermodynamic system. In the context of heavy ion collisions,
the coefficients of shear viscosity perhaps has been the mostly studied
transport coefficient. The spatial anisotropy in a nuclear collision
gets converted to a momentum anisotropy through a hydrodynamic evolution. 
The equilibration of momentum anisotropy is mainly controlled by shear
viscosity. Indeed, elliptic flow measurement at RHIC led to 
$\frac{\eta}{s}$, the ratio of shear viscosity ($\eta$) to the entropy
density $s$, close to $\frac{1}{4\pi}$ which is the smallest 
for any known liquid in nature \cite{hirano}. Indeed, arguments based on
ADS/CFT correspondence suggest that the ratio $\frac{\eta}{s}$ cannot be 
lower than this 'Kovtun-Son-Starinets' (KSS) bound \cite{kss}. Thus the 
quark gluon plasma(QGP) formed in the heavy ion collision is the 
most perfect fluid.

Apart from shear viscosity, the transport coefficient that relates the 
momentum flux with a velocity gradient is the bulk viscosity. 
Generally, it was
believed that the bulk viscosity does not play any significant role
in the hydrodynamic evolution of the matter produced in heavy ion collision experiments. The argument being that the bulk viscosity $\zeta$ scales like
$\epsilon-3 p$ and therefore will not play any significant role as the matter
might be  following the ideal gas equation of state. However, in course of
 the expansion of the fire ball the temperature can be near the critical temperature $T_c$ where $\epsilon-3p$ can be large as expected from the
lattice QCD simulations\cite{tanmoy,borsonyi} leading to large value of 
 bulk viscosity. This in turn can give rise to phenomena of cavitation when
the pressure vanishes and the hydrodynamic description for the evolution
breaks down\cite{cavitation}. 

 There has been various attempts to estimate bulk viscosity for strongly
interacting matter. The rise of bulk viscosity near the transition
temperature has been observed in various effective models of strong interaction.
These include chiral perturbation theory\cite{dobado}, quasi particle models
\cite{sasakiqp} as well as
Nambu-Jona-Lasinio model \cite{sasakinjl}. 
One of the interesting way to extract this
is using symmetry properties of QCD once one realizes that
the bulk viscosity characterizes the response to conformal transformation.
This was attempted in Ref.\cite{karschkharzeev}. Based on Kubo formula for the
$\zeta$ and the low energy theorems \cite{ellislet}
bulk viscosity gets related to
thermodynamic properties of strongly interacting system.

It may be noted that, it is also of both practical and fundamental importance to
know the transport coefficients in the hadron phase to distinguish the
 signatures of  QGP matter and hadronic matter. 
The computation of the transport coefficient of the hadronic mixture is not an easy task. There have been various attempt on this field over last few years
involving various approximations like relaxation time approximation, Chapman-Enscog as well as Green Kubo approach to estimate the shear viscosity to 
entropy ratio using different effective models for
hadronic interactions 
\cite{dobadoshear,sasakinjl,prakashwiranata,purnendu,greco}. 

In a different approach, $\frac{\eta}{s}$ has also been calculated within
a hadron resonance gas model in an excluded volume approximation \cite{gorenstein}
with a molecular kinetic theory approach to relate shear viscosity coefficient
to the average momentum transfer.  This was used later to include
the effects of rapidly rising hadronic density of states near the critical
temperature modeled by  hagedorn type exponential rise of density of states
\cite{greinerprl}. Such a description could describe the lattice data
and indicated that the hadronic matter could become almost a perfect fluid
where $\frac{\eta}{s}$ could approach the KKS bound. Later lattice data 
\cite{borsonyi} which indicated a lower pseudocritical temperature about 160 MeV led to the assertion that the hot hadronic matter described through hadron resonance gas is far from being a perfect fluid\cite{greinerprc}. All these studies
have been done at zero baryon density.

It has been also known that the basic features of hadronization
in heavy ion collisions are well described by the hadron resonance gas models. 
The multiplicities of particle abundances of various hadrons in
these experiments show good agreement with the corresponding thermal abundances
calculated in HRG model with appropriately chosen temperature
and chemical potentials \cite{hrgexp}. 
In the present work, we generalize the above approach of \cite{greinerprc}
for studying viscosity
coefficients within the ambit of hadron resonance gas model 
to include finite chemical potential effects. 
This can possibly have some relevance on the current and planned experiments
 with heavy ion collisions at {\it beam energy scan} at RHIC \cite{bes}, 
{\it compressed baryonic matter} at GSI \cite{cbm} and {\it neuclotron-based ion collider facility (NICA)}
at Dubna \cite{nica}. 

The shear viscosity to entropy ratio at finite baryon density has been
estimated using relativistic Boltzmann equations for pion nucleon system using
phenomenological scattering amplitude\cite{nakano,itakura}. This leads
to the ratio as a decreasing function of chemical potential in the T-$\mu$
plane. Further, this ratio as a function of chemical potential shows a
valley structure at low temperature which was interpreted as a signature of
liquid gas phase transition\cite{nakano,itakura}.

The bulk viscosity at finite chemical potential using low energy theorems 
of QCD has been studied in Ref.\cite{wang}. This was estimated using a Schwinger Dyson approach to calculate the dressed quark propagator at finite chemical
potential to use it for calculation of thermodynamical quantities needed
to estimate bulk viscosity. As mentioned, we shall estimate these viscosity
coefficients within the ambit of hadron resonance gas which can be a complimentary to the above approaches.

We organize the paper as follows. In the following section we recapitulate the
results of Ref.\cite{karschkharzeev} for bulk viscosity coefficient as related to the thermodynamic
quantities using Kubo formula and low energy theorems generalized to include finite chemical
potential terms. We also note down here the expression  for the shear viscosity using
quantum molecular dynamics method modified appropriately for relativistic system. In section III
we spell out the  the hadron resonance gas model including a hagedorn spectrum above a cut off and
the resulting thermodynamics. We estimate the quark condensates in a thermal, dense medium of hadron gas
in a subsection here. In section IV we discuss the results. Finally, we summarize and conclude in section V.

\section{Bulk and shear viscosity coefficients  at finite $T$ and $\mu$}
Bulk viscosity corresponds to the response of the system to conformal
transformations and can be written as per Kubo formula as a bilocal 
correlation function\cite{karschkharzeev}
\be
\zeta=\lim_{\omega\rightarrow 0}
\frac{1}{9\omega}\int_0^\infty d t\int d\zbf x \exp(i\omega t)
\left[\theta_\mu^\mu(x),\theta_\mu^\mu(0)\right]
\equiv\int d^4x \:iG^R(x)
\label{defbulk}
\ee

with $G^R(x)$ being the retarded function for the trace of energy momentum
tensor. One can introduce a spectral function $\rho(\omega,\zbf p)=
-(1/\pi)Im G(\omega,\zbf p)$ to write a dispersion relation for the
$G^R(\omega,\zbf p)$. Assuming an ansatz for the spectral function
at low energy\cite{karschkharzeev} as $\rho(\omega,\zbf 0)/\omega=
(9\zeta/\pi)(\omega_0^2/(\omega_0^2+\omega^2)$, where, $\omega_0$ is a scale
at which perturbation theory becomes valid , the bulk viscosity
can be written as
\be
9\zeta\omega_0=2\int_0^\infty du \frac{\rho(u,0)}{u}du=
\int
d^4x \langle \theta_\mu^\mu(x)\theta_\mu^\mu(0)\rangle\equiv\Pi
\label{pi}
\ee
The stress energy tensor for QCD is given as
\be
\theta_\mu^\mu=m\bar qq+\frac{\beta(g)}{2g}G^a_{\mu\nu}G^{a\mu\nu}\equiv
\theta_q+\theta_g
\ee
In the above $g$ is the strong coupling and $\beta(g)$ is the QCD beta
function that decides the running of the QCD coupling. Thus the 
evaluation of the bulk viscosity reduces to evaluation of the
stress energy correlator. This is done by using the low energy theorems
of QCD generalized to finite temperature and density according to which
for any operator $\hat O$, its correlator with the gluonic part of the
stress tensor $\theta_g$ is given as
\be
\int d^4x \langle \theta_g(x)\hat O)\rangle=(\hat D-d)\langle\hat O
\rangle(T,\mu),
\label{lettmu}
\ee
 where, $\hat D =T\partial/\partial T+\mu\partial/\partial \mu-d$, with
$d$ being the canonical dimension of the operator $\hat O$. 
Using Eq.(\ref{lettmu}) in Eq.(\ref{pi}) one has
\bearr
\Pi &=& (\hat D-4)\langle\theta_\mu^\mu\rangle + (\hat D-2)\langle{\theta^q}^\mu_\mu\rangle\nonumber\\
&=& 16|\epsilon_{vac}^g|+6 (f_\pi^2m_\pi^2+f_k^2m_k^2)\nonumber\\
&+&TS(\frac{1}{c_s^2}-3)+(\mu\frac{\partial}{\partial\mu}-4)(\epsilon^*-3p^*)+(\hat D-2) m_q\langle\bar qq\rangle_{*}
\label{pi1}
\eearr

In the above we have used $\langle\theta_\mu^\mu\rangle=\epsilon-3 p$ and the thermodynamic 
relations $c_v=\partial\epsilon/\partial T$,
$\partial p/\partial T=s$ and $c_s^2=S/c_v$ for the velocity of sound of
 the medium. We have also separated the
contributions to the correlators in terms of the vacuum and the medium. In Eq.(\ref{pi1}) we have neglected 
terms quadratic in the current quark masses and have used PCAC relations to express vacuum condensates to the
masses and decay widths of pions and kaons. It is trivial to check that
for $\mu=0$ Eq.(\ref{pi1}) reduces to the main results of 
Ref.\cite{karschkharzeev}. For T=0 and $\mu\neq 0$, one can simplify Eq.(\ref{pi1})and Eq.(\ref{pi}) reduces to
\be
9\zeta(\mu)\omega_0=16P(\mu)-7\mu\rho+\mu^2\frac{\partial\rho}{\partial\mu}
+(\mu\frac{\partial}{\partial\mu}-2)m\langle\bar qq\rangle
\label{zetamu}
\ee
We might note here that the above expression differs from the same given in
Ref.\cite{wang}.This, however, matches with the expression 
given in Ref.\cite{agasian} in
the appropriate limit,
where, bulk viscosity was computed including the effects of magnetic field
at finite baryon densities and temperature.

Thus the coefficients of bulk viscosity gets related to the vacuum properties of QCD as well as to the  equilibrium
thermodynamic system parameters of QCD like the velocity of sound, non-ideality and the in medium quark condensates. These
thermodynamic quantities shall be estimated within hadron resonance gas model which we shall spell out in the next
section.

Next, we consider the shear viscosity coefficient $\eta$ for the hadronic 
medium. It is known that hadrons interact in various channels and there
 is possibility of attractive and repulsive interactions. Within the
 hadron resonance model, the attractive channels are effectively
 included by including the resonances and the repulsive channels can 
be modeled in a simple manner through and excluded volume 
correction \cite{hagedorn,kapustaolive,rischkegorenstein}. The shear
viscosity in a relativistic gas of multi component hard core spheres
can be written as \cite{gorenstein,greinerprc}
\be
\eta=\frac{5}{64\sqrt 8 r^2}\sum_i\langle|\zbf p|\rangle \frac{n_i}{n}
\label{eta}
\ee
where, $\langle|\zbf p|\rangle$ is the average momentum of the i-th species 
particles and $r$ corresponds to hard core radius of each hadron .
Further, in the above, $n_i$ is the
number density of the i-th particle species and $n=\sum_in_i$.
\section{Hadron Resonance Gas model}
The central quantity in the hadron resonance gas models (HRGM) is the thermodynamic potential which is that of
a free boson or fermion gas and is given as
\be
\log(Z,\beta,\mu,)=\int dm \left(\rho_M(m)\log Z_b(m,v,\beta,\mu)+\rho_B(m)\log Z_f(m,v,\beta,\mu)\right)
\label{logz}
\ee
where, the gas of hadrons is contained in a volume V, at a temperature $\beta^{-1}$ and chemical potential
$\mu$. $Z_b$, $Z_f$ are the partition functions of boson and fermions respectively with mass $m$. Further, $\rho_M$ and $\rho_B$ are the
spectral densities of bosons and fermions respectively. Using Eq.(\ref{logz}), 
one can calculate the energy density $\epsilon$ by taking derivative 
with respect to $\beta$, pressure p, by taking a derivative with 
respect to $V$, number density $\rho$ by taking a derivative with
 respect to $\mu$. One can also find out the trace anomaly $\epsilon-3p$,
 entropy density, specific heat as well as the speed of sound from these
 quantities. 

Hadron properties enter in these models through the spectral densities $\rho_{B/M}(m)$. One common approach in 
HRGMs is taking all the hadrons and their resonances up to a mass cutoff $\Lambda$ and write
\be
\rho_{B/M}(m)=\sum_{i}^{M_i<\Lambda} g_i\delta(m-M_i)
\label{hrgm1}
\ee
where, the sum is over all the baryons or meson states up to a mass that is
less than the cut off $\Lambda$. $M_i$ are the masses of the known 
hadrons and  $g_i$ is the degeneracy factor (spin, isospin).
On the other hand, an exponentially increasing density of state 
was necessary to explain the rapid increase in entropy density near
the transition region in lattice QCD simulation \cite{majumdermueller}. Such exponential rise of density of states has also been used to study
observables like dilepton production \cite{leonidov}
 as well as chemical equilibration\cite{igorgreiner}. Motivated by
 such observations we take the modified spectral function as\cite{cleymans,guptagod,greinerprl}
\be
\rho_{B/M}(m)=\sum_{i}^{M_i<\Lambda} g_i\delta(m-M_i)+\rho_{HS}(m)
\label{hrgm2}
\ee
where $\rho_{HS}(m)$ is the spectral density for the heavier Hagedorn states(HS).
To describe the much needed large density of states, one can take an exponentially rising density of state \cite{Hagedorn}
for $\rho_{HS}$ beyond the cut-off $\Lambda$ 
which implies an underlying string picture for hadrons. On the other hand, 
one can also consider a simple power law form introduced in 
Ref.\cite{shuryak}
as a nice alternative to describe the rise of the hadronic mass 
spectrum \cite{shuryak}.   
We shall consider here both the forms for the continuum part of the spectral density given as
\be
\rho_{exp}=\frac{A}{({m^{2}+m_{0}^{2}})^{2}}e^{\frac{m}{T_{H}}}
\label{rhoexp}
\ee
\be
\rho_{power}=\frac{A}{T_{H}}\bigg(\frac{m}{T_{H}}\bigg)^{\alpha}
\label{rhopower}
\ee
where parametrization of the two spectral forms is given in table below.

\begin{center}
    \begin{tabular}{ | l | l | l |l |p{0.5cm} |}
    \hline
    spectral density & $T_{H}(GeV)$ & \ A & $m_{0}(GeV)$ & \ $\alpha$\\ \hline
    \ \ \ \ \ \ \ $\rho_{exp}$ & \ \ \ 0.210 & 0.63 & \ \ \ \ 0.5 & \ -\\ \hline 
    \ \ \ \ \ \ \ $\rho_{power}$  & \ \ \ 0.180 & 0.51 & \ \ \ \ -& \  3 \\ \hline
   
    \end{tabular}
 
\end{center}

 We have taken the parameters $A$ and $m_0$ for $\rho_{exp}$ as in Ref.
\cite{majumdermueller} and taken a different value for $T_H$ so as to 
fit the lattice data of Ref.\cite{borsonyimu}.  Similarly the parameters 
$\alpha$ and $T_H$ for $\rho_{power}$ is taken so as to fit the lattice
data of Ref.\cite{borsonyimu} while keeping the parameter $A$ same as taken in 
Ref.\cite{greinerprc}.

With the ansatz for the spectral densities, the pressure $P=P_M+P_B$ arising 
from mesons and baryons respectively are given by
\bearr
P_M &=&\frac{1}{2\pi^2}\bigg[-\sum_ig_i\int k^2 dk\log\left(1-\exp(-\beta\epsilon_i)\right)\nonumber\\
&+&\int_\Lambda^\infty \rho_{HS}(m) dm \frac{m^2}{\beta^2}K_2(\beta m)\bigg]
\label{pmes}
\eearr
\bearr
P_B &=&\frac{1}{2\pi^2}\bigg[-\sum_ig_i\int k^2 dk\bigg(\log\big(1-\exp(-\beta(\epsilon_i-\mu))\big)
+\log\left(1-\exp(-\beta(\epsilon_i+\mu))\right)\bigg)\nonumber\\
&+& 2\int_\Lambda^\infty \rho_H(m) dm \frac{m^2}{\beta^2}K_2(\beta m)\cosh(\beta\mu)\bigg ]
\label{pbar}
\eearr
Here, $K_n(x)$ is the modified Bessel function of order $n$.
Similarly, the energy density $\epsilon=-\frac{1}{\beta}
\frac{\partial}{\partial\beta}(\beta p)+\mu\frac{\partial}{\partial \mu}p=\epsilon_M+\epsilon_B$ ,
with the energy density
 of mesons $\epsilon_M$ given as
\bearr
\epsilon_M &=&
\frac{1}{2\pi^2}\bigg[\sum_ig_i\int k^2dk 
\frac{\epsilon_i}{\exp(\beta\epsilon_i)-1} 
\nonumber\\
&+&\int_\Lambda^\infty \rho_{HS}(m)dm m^4
\left(\frac{3}{\beta^2m^2}K_2(\beta m)
+\frac{1}{\beta m}K_1(\beta m)\right)\bigg]
\label{emes}
\eearr
and, the contribution of the baryons to the energy density $\epsilon_B$ is given as
\bearr
\epsilon_B& =&
\frac{1}{2\pi^2}\bigg[\sum_ig_i\int k^2dk {\epsilon_i}
\left(\frac{1}{\exp(\beta(\epsilon_i-\mu))+1} 
+\frac{1}{\exp(\beta(\epsilon_i+\mu))+1}\right)\nonumber\\
& +&
\int_\Lambda^\infty \rho_H(m)dm m^4\left(\frac{3}{\beta^2m^2}K_2(\beta m)
+\frac{1}{\beta m}K_1(\beta m)\right)\bigg]
\label{ebar}
\eearr
The baryon number density is given by
\bearr
n_B&=&\frac{1}{2\pi^2}\bigg[{g_i}\int k^2dk \left(
\frac{1}{\exp(\beta(\epsilon_i-\mu))+1} -
\frac{1}{\exp(\beta(\epsilon_i+\mu))+1}\right)\nonumber\\
&+&
2\int_\Lambda^\infty \rho_H(m)dm \frac{m^2}{\beta^2}K_2(\beta m)\bigg]
\label{rhob}
\eearr

Using these quantities one can calculate the other quantities like the interaction measure $\epsilon-3p$, 
entropy density $s=(\frac{\partial p}{\partial T})$ as needed for the estimation of bulk viscosity.
\subsection{quark condensates in the hadronic medium}
 The other quantity we need to know is the quark condensates in the medium 
to estimate the bulk viscosity. To estimate this within the
framework of HRGM, it is necessary to know the dependence of hadron masses on the current quark masses. The chiral condensate is 
given in terms of the thermodynamic potential (negative of the pressure)  as 
$\langle\bar q q\rangle=-\frac{\partial p}{\partial m_q}$ which leads to
\be
\langle\bar q q\rangle=\langle\bar q q\rangle_0+\sum_{mesons}\frac{\sigma^M}{m_q }n_M
+\sum_{baryons}\frac{\sigma^B}{m_q }n_B
\label{mediumcondensate}
\ee
where $n_M$ and $n_B$ are the scalar densities of mesons and baryons given respectively as
\be
n_M=\frac{g_i}{2\pi^2}\int k^2 dk \frac{m_M}{\epsilon_M}\frac{1}{\exp(\beta\epsilon_M)-1},
\ee
\be
n_B=\frac{g_i}{2\pi^2}\int k^2 dk \frac{m_B}{\epsilon_B}\left(\frac{1}{\exp(\beta(\epsilon_B-\mu_B))+1} +\frac{1}{\exp(\beta(\epsilon_B+\mu_B))+1}\right)
\ee
 Further, the $\sigma^{M/B}$ is the hadronic sigma term i.e. the response of hadronic masses to the changes of the current quark masses
\be\sigma_q^{M/B}=m_q\frac{\partial M_{M/B}}{\partial m_q}
\ee
Thus computing the behavior of in-medium condensate  within HRGM reduces to the problem of calculating the $\sigma$-terms of the hadrons.
We do this in a manner similar to given in Ref.\cite{blaschke}. For the pseudoscalar bosons, we use the Gell Mann-Oakes-Renner (GOR) relation
to have
\be
\frac{\partial m_\pi^2}{\partial m_q}=-\frac{\langle\bar q q\rangle_0}{f_\pi^2}\left(1+2\kappa \frac{m_pi^2}{f_\pi^2}\right)
\ee
\be
\frac{\partial m_K^2}{\partial m_{q,s}}=-\frac{\langle\bar q q\rangle_0+\langle\bar s s\rangle_0
}{2f_K^2}\left(1+2\kappa \frac{m_K^2}{f_\pi^2}\right)
\ee
with the parameter $\kappa=0.021\pm0.008$ \cite{jaminplb}. here, we have taken $m_q=m_u=m_d=5.5 MeV$, $m_S=138 MeV$, $f_\pi=92.4MeV$,
$f_K=113 MeV$$\langle\bar u u\rangle_0=\langle\bar d d\rangle_0=\langle\bar q q\rangle_0= (-240 MeV)^3$, $\langle\bar s s\rangle_0=0.8
\langle \bar q q\rangle_0$.
For the other hadrons we use a model based on valence quark 
structure as in Ref. \cite{leupold}. Here the masses of the baryons (B) or mesons
(M) scale as
\be
m_B=(3-N_s) M_q+N_sM_s +\kappa_B
\ee
\be
m_M=(2-n_s)M_q+N_sM_s+\kappa_m
\ee
In the above, $M_q$, $M_s$ are the constituent quark masses for the light and strange quarks respectively, $\kappa_{B/M}$ are
constants depending upon the hadronic state but not on current quark masses and $N_s$ is the numbers of strange quarks. The constituent quarks  $M_q$ and $M_s$ partially account for the strong interaction dynamics. For computation of the $\sigma$ term one meeds to know the variation
of the constituent quarks with current quark masses. This dependence is taken 
from Nambu-Jona-Lasinio model \cite{hmnjl} where, the dynamical
mass changes by 14 MeV as the current quark mass is changed from 0 to 5.5 MeV. Similarly for strange quark the mass change is about
235.5 MeV as current quark mass is varied from 0 to 140.7 MeV. 
This e.g. results in $\sigma$ terms for nucleons and $\Lambda$ hyperon 
as 42 MeV and 263.5 MeV respectively. 
\section{Results and discussions}
Let us first discuss the thermodynamics of hadron resonance gas
specified by the spectral density as given by Eq.(\ref{hrgm2}).
To estimate different thermodynamic quantities,  for the discrete
part of spectrum , we have taken all the hadrons and their
 resonances with mass less than 2 GeV \cite{pdgb}.
For the Hagedorn part, consider both the forms of spectral density 
given by Eq. (\ref{rhoexp}) and Eq.(\ref{rhopower}).

\begin{figure}[h]
\vspace{-0.4cm}
\begin{center}
\begin{tabular}{c c}
 \includegraphics[width=9cm,height=7cm]{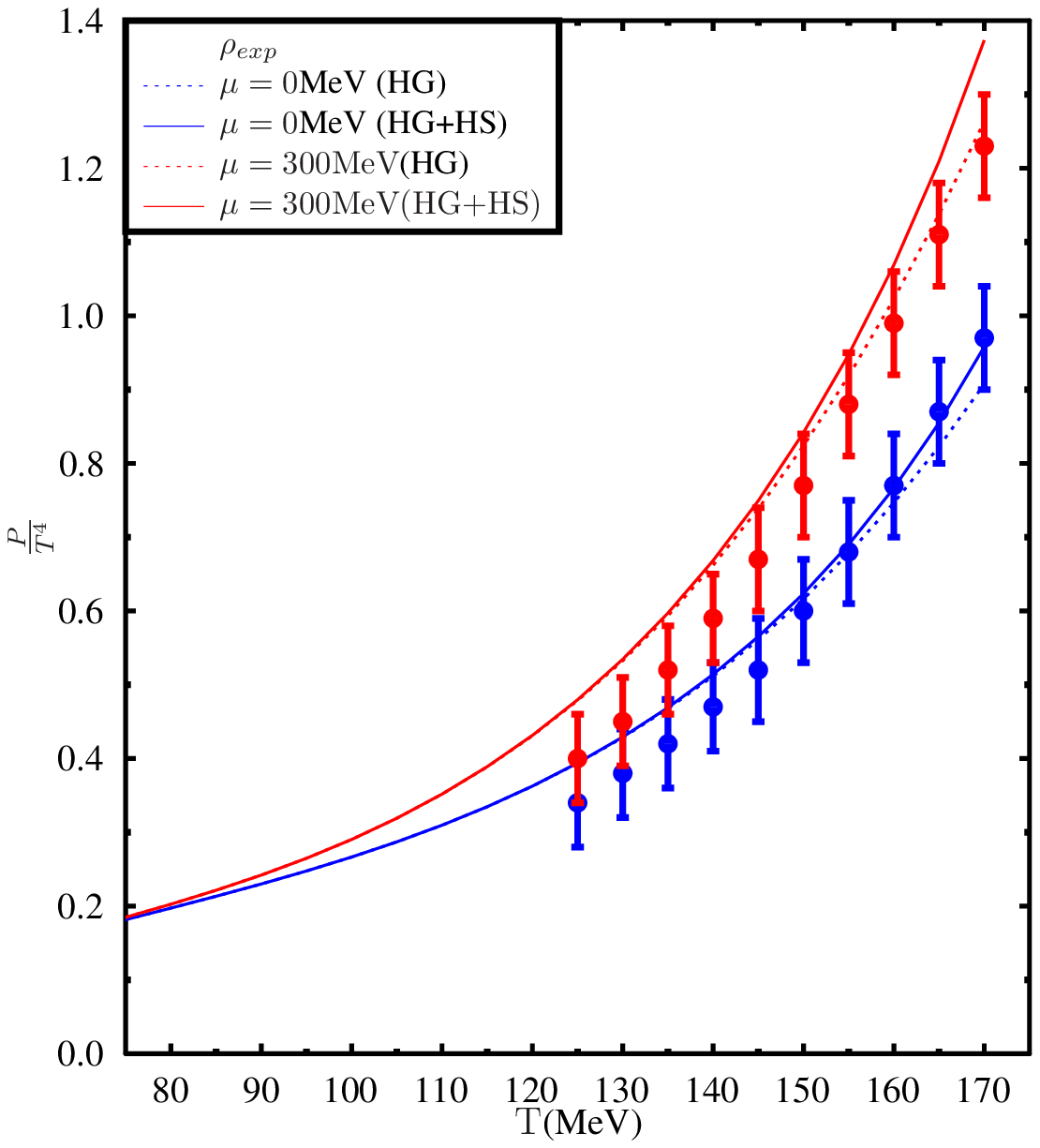}&
  \includegraphics[width=9cm,height=7cm]{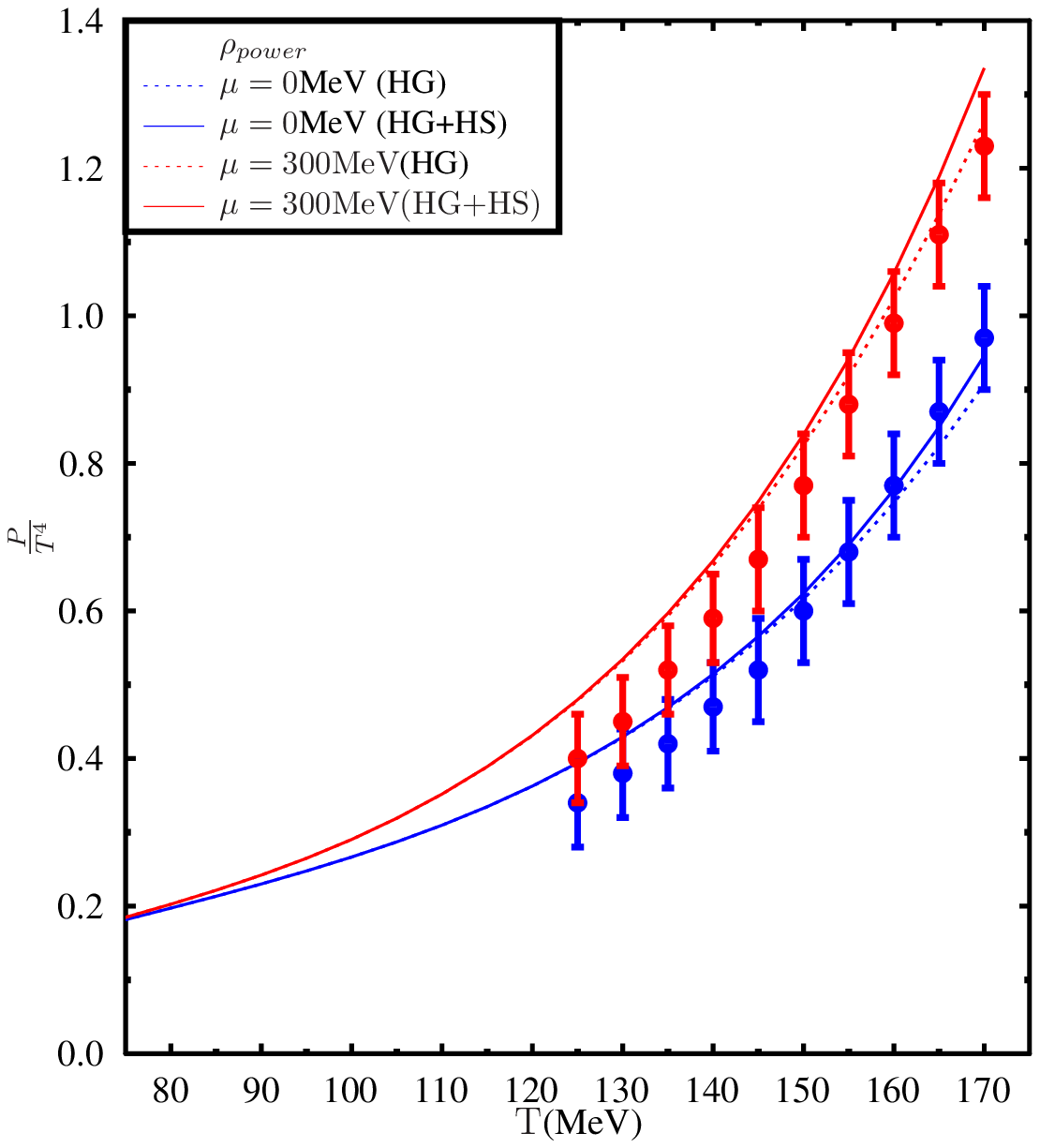}\\
 (1 a)&(1 b)
  \end{tabular}
  
\caption{ Thermodynamics of   hadron resonance gas.  Left panel (Fig. 1 a)   shows
pressure as a function of temperature for  $\mu_b=0$(blue) and $\mu_B=300$ MeV (red) with the hagedorn  spectrum $\rho=\rho_{exp}$ as
in Eq.(\ref{rhoexp}). The dotted line correspond to taking discrete spectrum for hadron resonance gas. The right panel
shows the same quantities but with the spectral function $\rho=\rho_{power}$ as given in Eq.(\ref{rhopower}). The data points
are from the lattice simulation results taken from Ref. \cite{borsonyimu}.
}\label{thermo1}
 \end{center}
 \end{figure}

 In Fig.\ref{thermo1} we have plotted the pressure in units of $T^{4}$ 
for two different chemical potentials, $\mu=0$ MeV and $\mu=300$ MeV. 
The lattice points with the error bars have been taken from the
 table 4 of the Ref. \cite{borsonyimu} corresponding to the 
continuum extrapolation. The dotted lines in Fig.\ref{thermo1}
 correspond to considering only the
discrete part of the spectral density in Eq.(\ref{hrgm1}) . Left panel correspond to exponential form of spectral density for continuum part while right
 panel corresponds to power law form of spectral density in Eq.(\ref{hrgm2}). 
As can be noted in this figure, the discrete spectrum coupled with 
continuum spectrum describe the lattice data quite well up to $T=170 MeV$ 
with the parametrization given in table 1 within the error bars of 
the lattice simulations.
 
\begin{figure}[h]
\vspace{-0.4cm}
\begin{center}
\begin{tabular}{c c}
 \includegraphics[width=9cm,height=7cm]{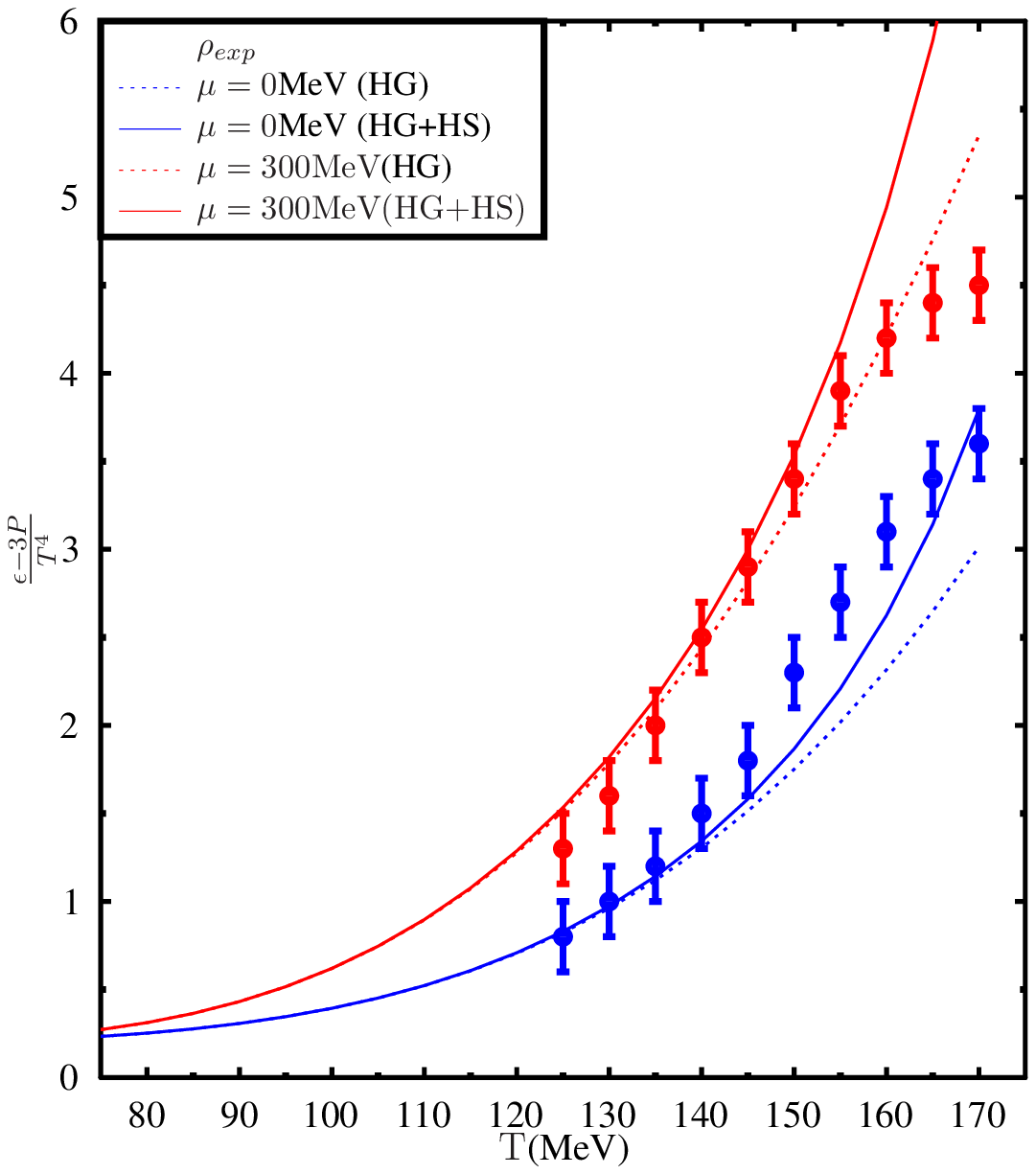}&
  \includegraphics[width=9cm,height=7cm]{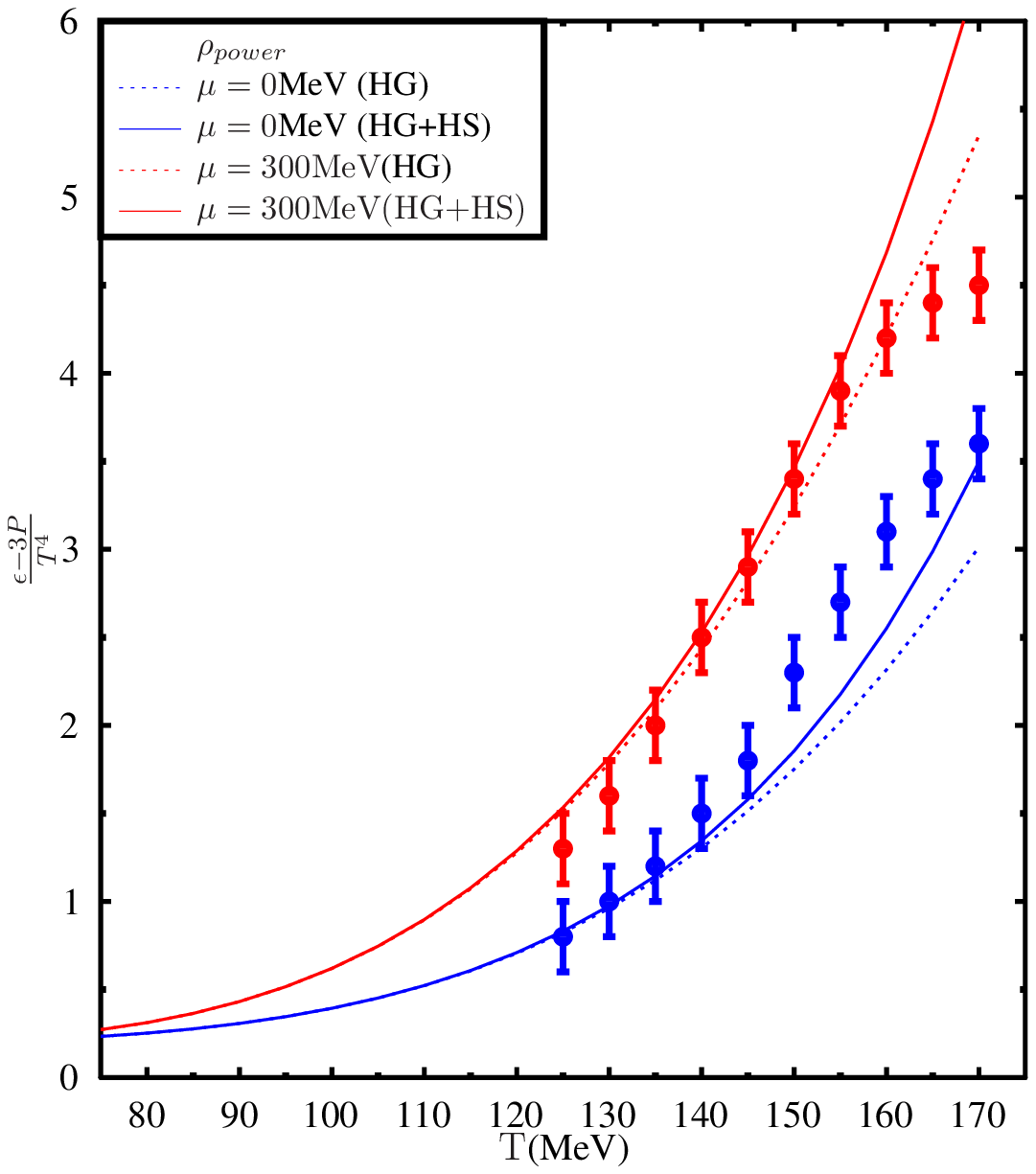}\\
 (2 a)&(2 b)
  \end{tabular}
\caption{Scale anomaly as a function of temperature for exponential spectral density (2 a) and power law spectral density
function (2 b). }
\label{thermo2}
\end{center}
 \end{figure}

 In fig. \ref{thermo2} we have plotted the dimensionless scale anomaly 
$(\epsilon-3p)/T^{4}$ as a function of temperature at two different 
chemical potentials. As can be noted from both the Fig.s (2a) and (2b), 
the discrete part of the spectral density does not give a good fit to the
lattice data beyond $140 MeV$, but when coupled with continuum part 
as in Eq.(\ref{hrgm2}) gives good fit to lattice data up to $150 MeV$ 
even at $\mu=300 MeV$ reasonably well. We have taken a higher $T_{H}$ 
value compared to \cite{greinerprc} that was required to fit the lattice 
data\cite{borsonyimu}. This is because, in Ref.\cite{greinerprc},
the lattice data was taken for $N_t=10$ lattice data of Ref.\cite{borsonyi}
while we have fitted with the continuum extrapolation of for $\mu=0$
 the lattice data in Ref.\cite{borsonyimu}. 

\begin{figure}[h]
\vspace{-0.4cm}
\begin{center}
\begin{tabular}{c c}
 \includegraphics[width=9cm,height=7cm]{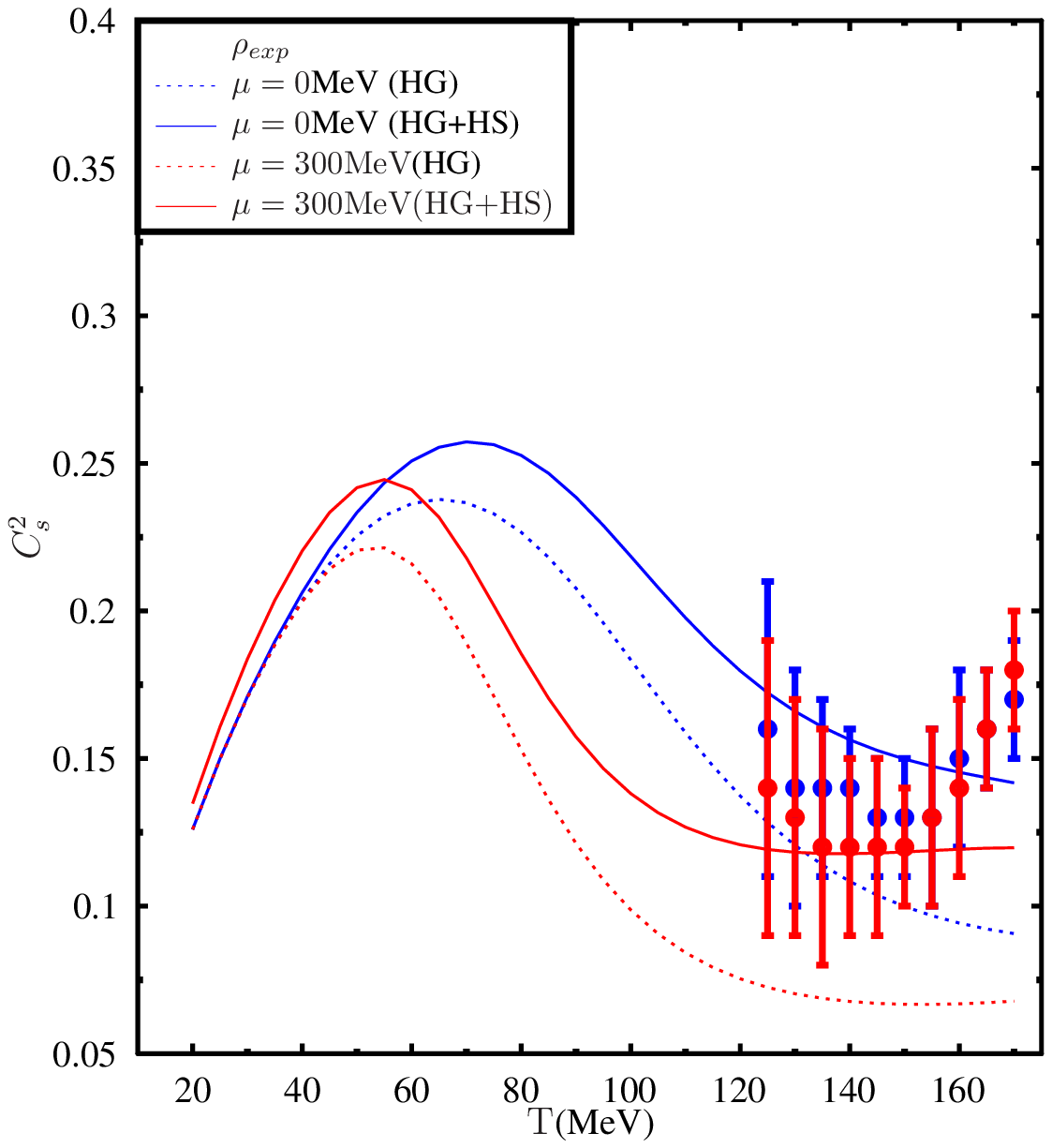}&
  \includegraphics[width=9cm,height=7cm]{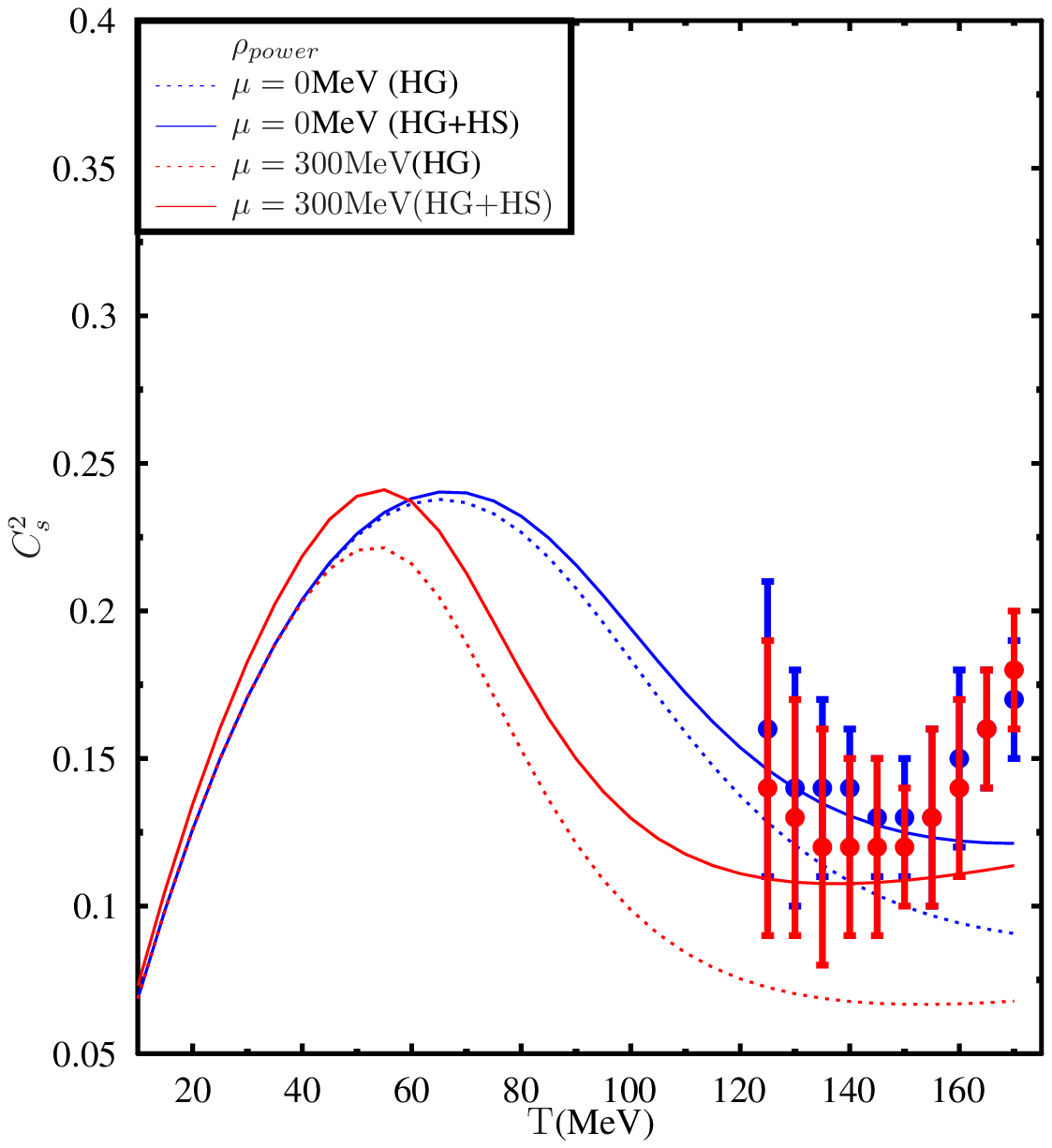}\\
 (3 a)&(3 b)
  \end{tabular}
\caption{Square of sound velocity as a function of temperature for exponential spectral density (3 a) and power law spectral density
function (3 b). }
\label{thermo3}
\end{center}
 \end{figure}
Fig \ref{thermo3} shows speed of sound squared ($C_{s}^{2}$) as a 
function of temperature at fixed values of chemical potential along with the
lattice simulation results of Ref.\cite{borsonyimu}.
 As can be noted from the figure,  keeping only the
 discrete part of the spectral density,
does not fit the lattice results although the same could fit the
lattice result for pressure and the scale anomaly results 
of Ref.\cite{borsonyimu}. On the other 
hand the power law parametrization for the continuum part of spectral
density  along with the discrete part leads to a reasonable fit to
lattice data up to $150 MeV$ both at $\mu=0$ and $\mu=300MeV$. The initial rise
in sound velocity with temperature is reflection of the fact that the
light degrees
of freedom are excited easily at low temperature and contribute to
pressure and energy. But at larger temperatures
when baryons are excited,
they
 contribute significantly to energy density but almost nothing to pressure.
 This leads to  decrease of  sound velocity with temperature seen
 at  higher temperatures ($T> 80 $MeV).  As chemical potential
increases, heavier baryonic channels opens up at low temperature 
and contribute to energy density significantly but nothing to pressure.
 This leads to  lower values of $C_{s}^{2}$ as the chemical potential
is increased.


\begin{figure}[h]
\vspace{-0.4cm}
\begin{center}
\begin{tabular}{c c}
 \includegraphics[width=9cm,height=7cm]{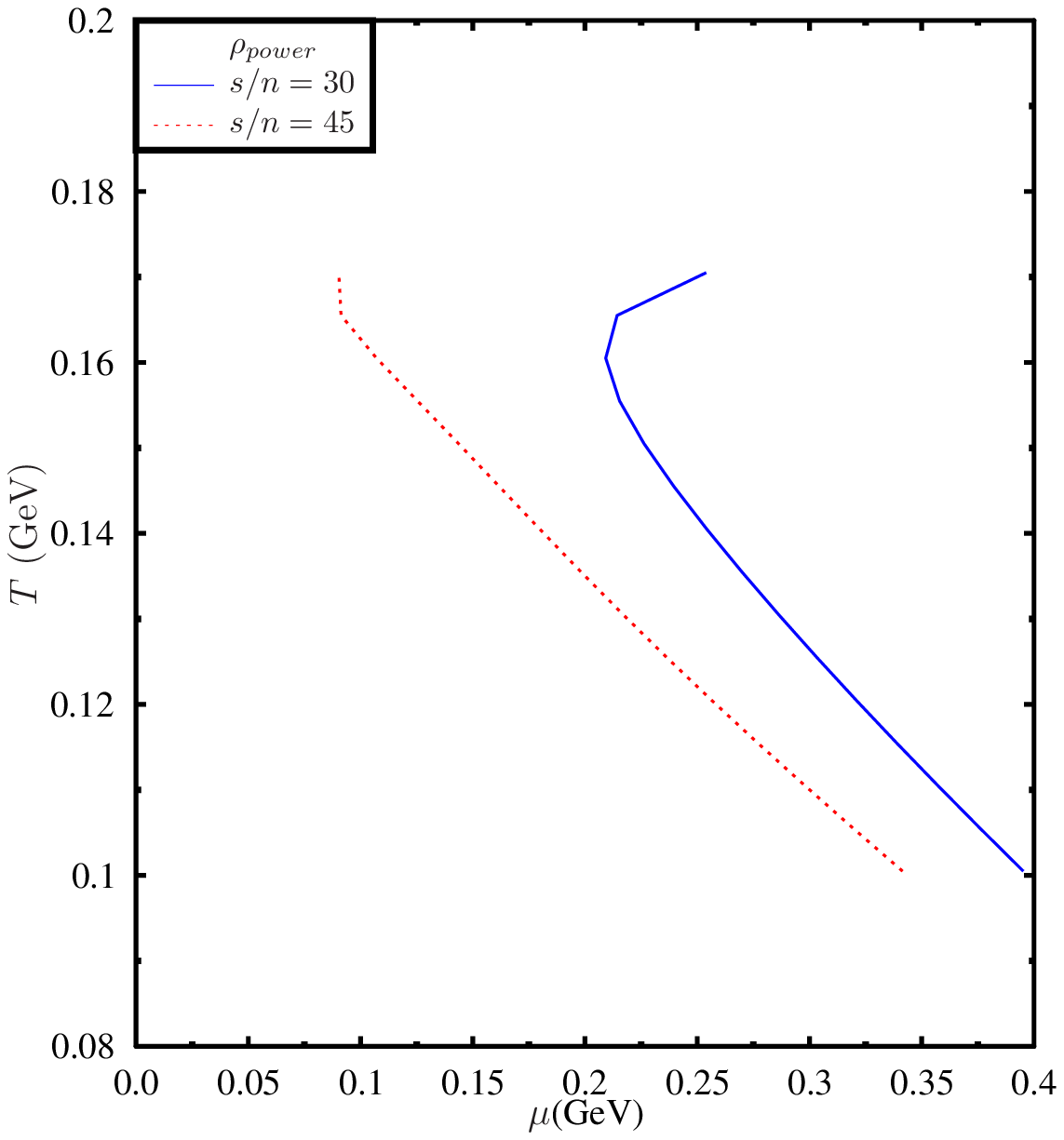}&
  \includegraphics[width=9cm,height=7cm]{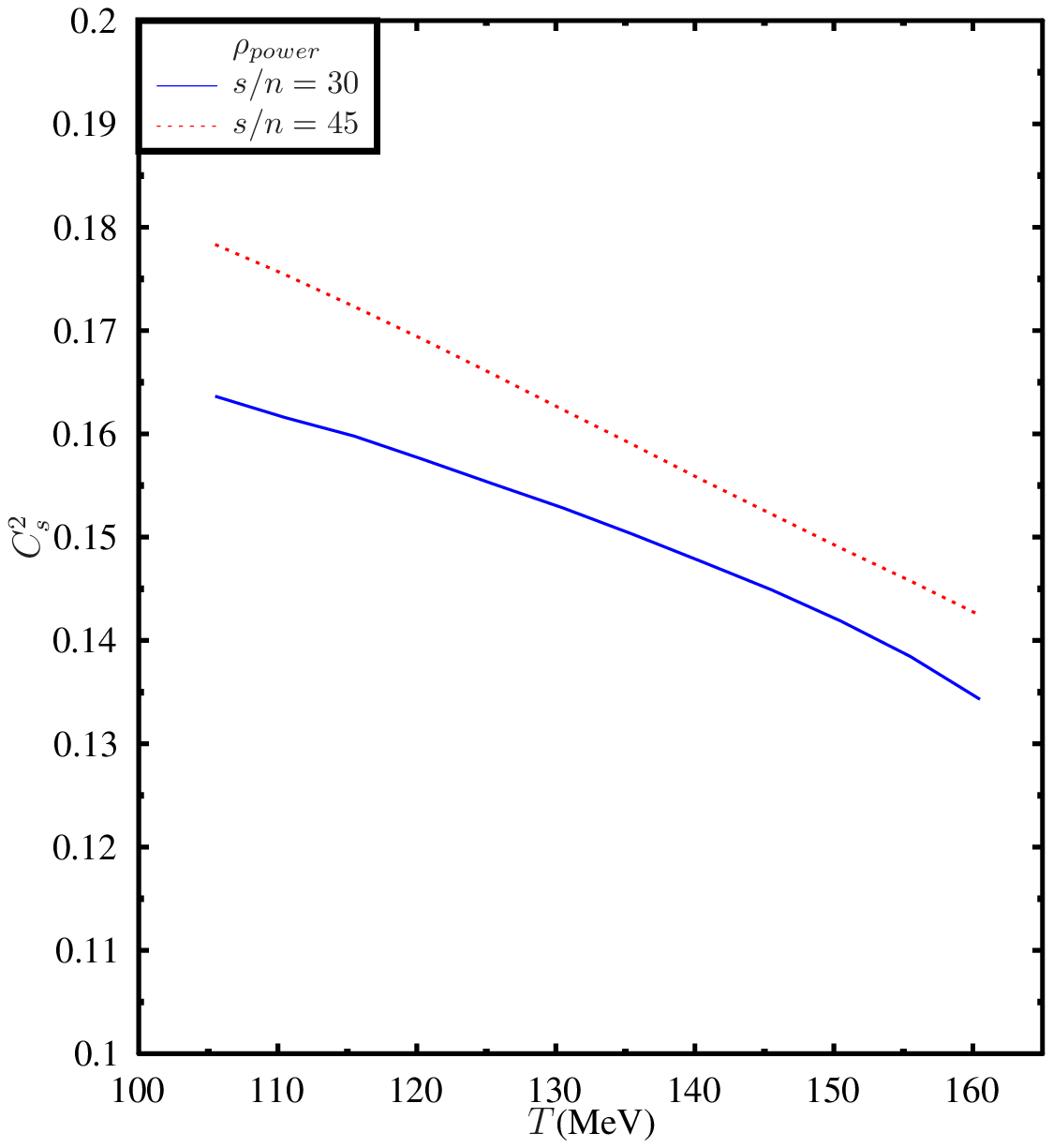}\\
 (4 a)&(4 b)
  \end{tabular}
\caption{   Velocity of sound at constant entropy per baryon rations. Left panel (4a) shows trajectories of constant
entropy per baryon in the phase diagram. velocity of sound for constant entropy per baryon is
plotted in Fig. 4b. }
\label{isentropic}
\end{center}
 \end{figure}

We have  also plotted speed of sound for isentropic situation in figure 
\ref{isentropic}.  To get the chemical
potential for a given temperature, we vary chemical potential 
so that ratio $S/N$ is constant. Resulting isentropic trajectories in
 the $\mu-T$ phase space is shown in fig. (4a). $S/N=30$ and $S/N=45$ 
corresponds to AGS and SPS \cite{blum}. 
As expected from the results for constant chemical potential
 (Fig.\ref{thermo3}), sound velocity is lower for lower $S/N$.

\begin{figure}[h]
\vspace{-0.4cm}
\begin{center}
\begin{tabular}{c c}
 \includegraphics[width=9cm,height=7cm]{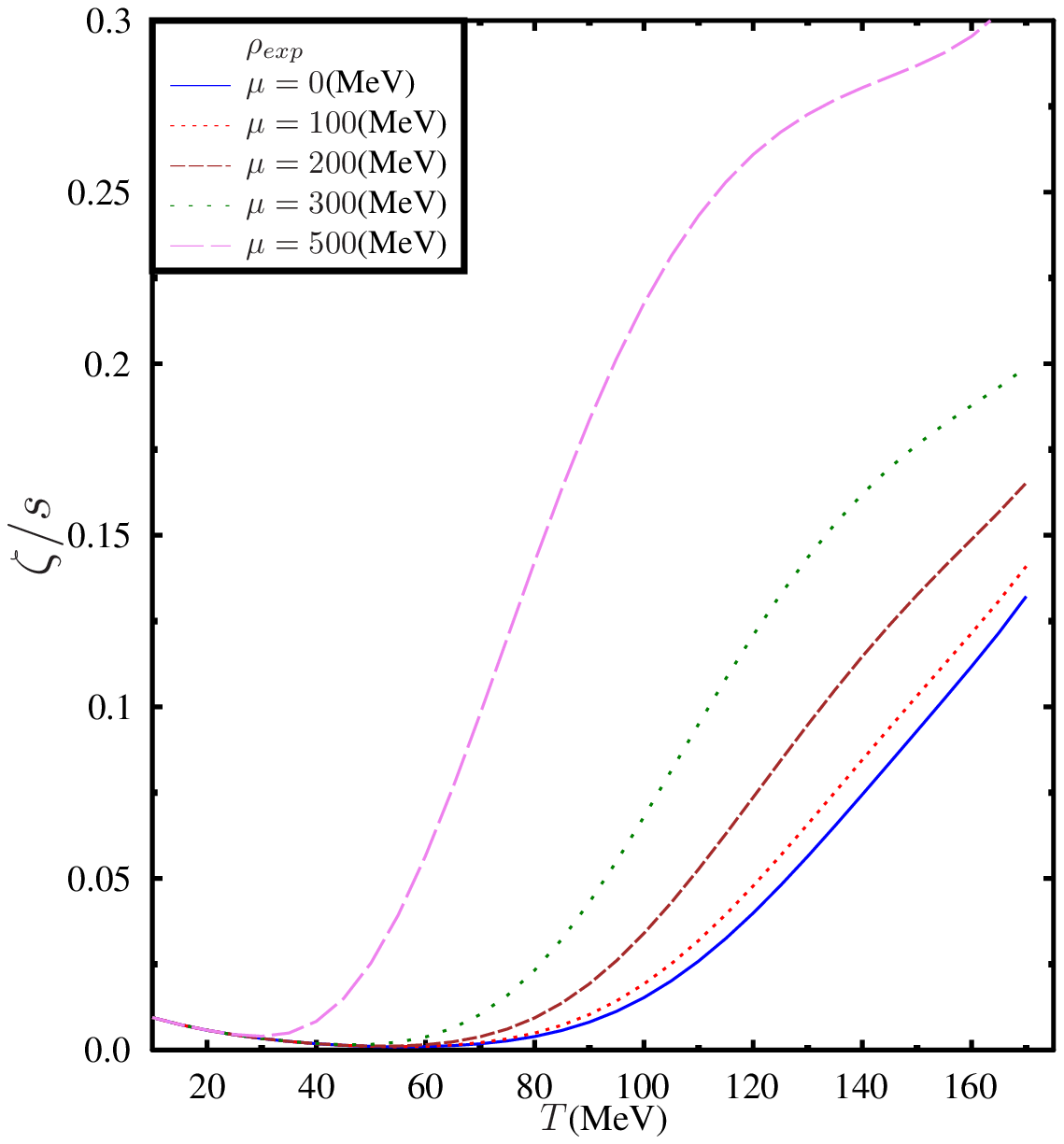}&
  \includegraphics[width=9cm,height=7cm]{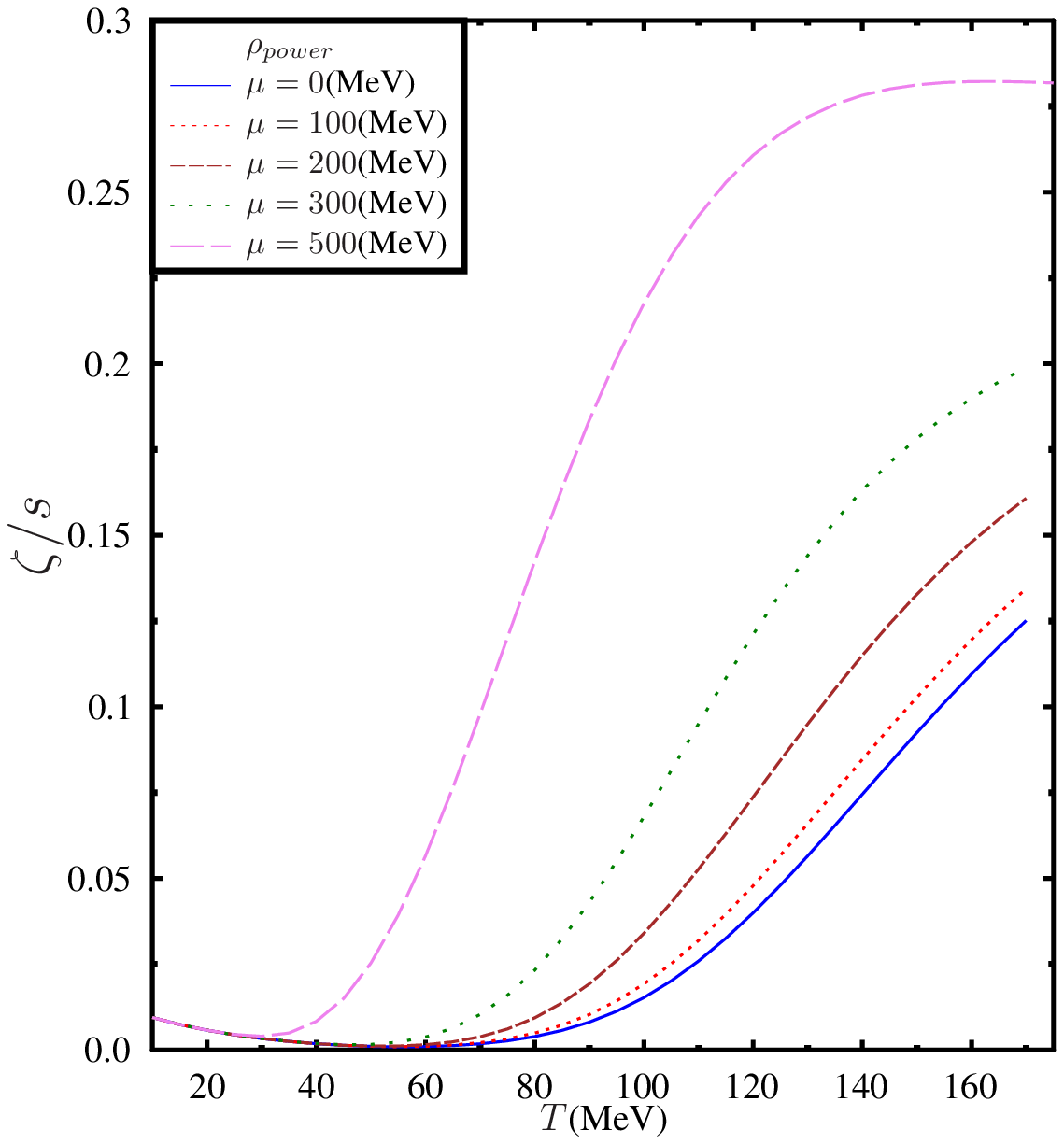}\\
 (5 a)&(5 b)
  \end{tabular}
  \caption{ Bulk viscosity to entropy ratio as a function of temperature for different chemical potentials. Left panel is with exponential hagedorn spectrum
and the right panel is with power law hagedorn spectrum} 
\label{zetabis}
  \end{center}
 \end{figure}

\begin{figure}[h]
\vspace{-0.4cm}
\begin{center}
\begin{tabular}{c c}
 \includegraphics[width=9cm,height=7cm]{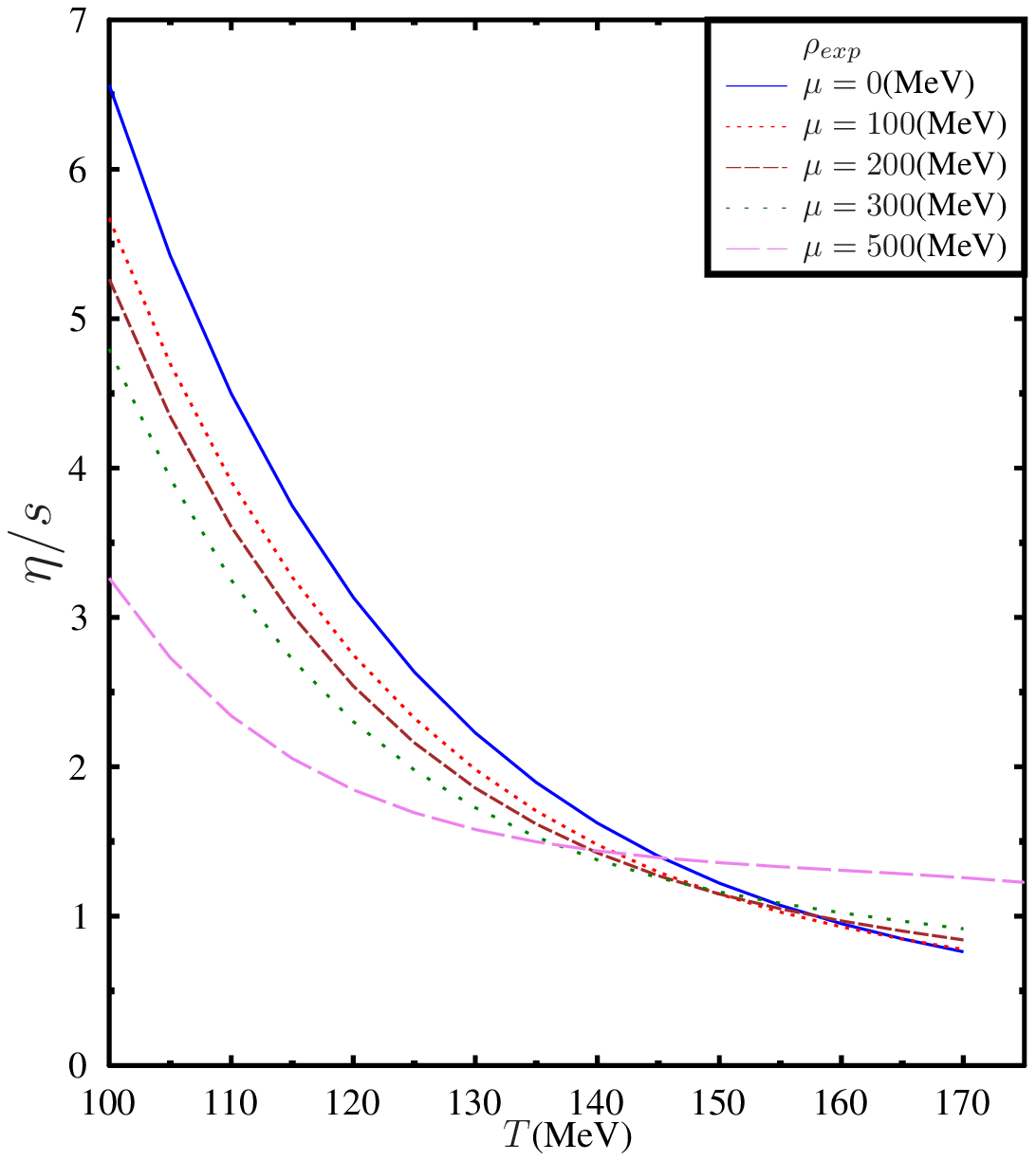}&
  \includegraphics[width=9cm,height=7cm]{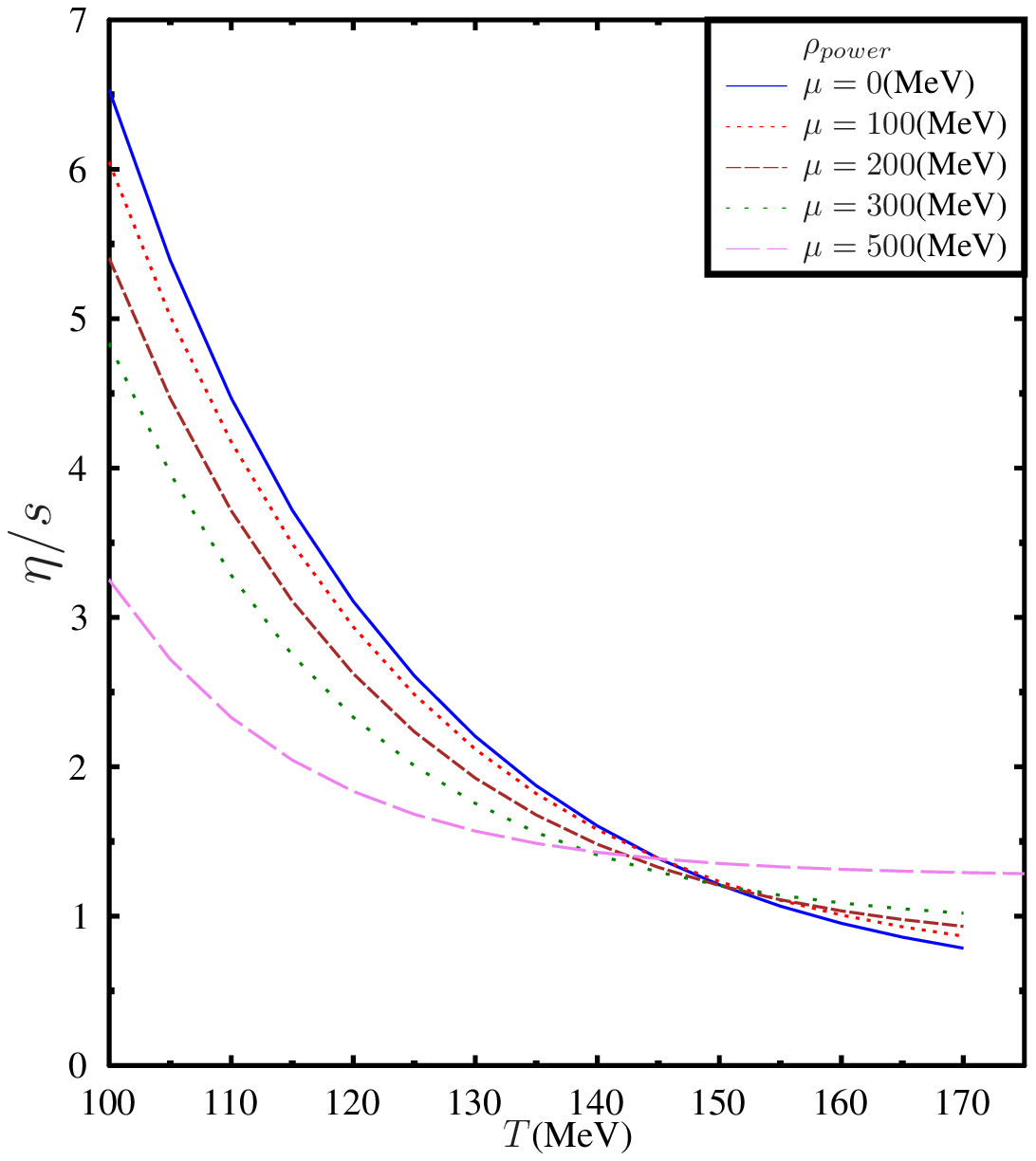}\\
 (6 a)&(6 b)
  \end{tabular}
\caption{   Shear viscosity to entropy ratio in the hadronic phase. Left 
panel (6 a) shows $\frac{\eta}{s}$ as a function
of temperature for different chemical potential with the exponential hagedorn spectrum . The right panel shows the same  with the power law hagedorn spectrum.
}
\label{etabis}
\end{center}
 \end{figure}

We next use these thermodynamic results for the hadron resonance gas 
to Eq.(\ref{pi}) and Eq.(\ref{pi1}) to estimate the bulk viscosity. We also
include here the contributions from the quark condensates in the discrete
part of the spectrum using Eq.(\ref{mediumcondensate}). Contribution of these
terms  to $\zeta/s$ turns out to be only few percent of the the total
 contribution.
The resulting behavior of $\zeta/s$ 
as a function of temperature is shown in Fig.\ref{zetabis} for different 
values of the baryon chemical potential. In general, the ratio
decrease with temperature at low temperature followed by a sharp increase
and finally flattens out at temperatures around 160 MeV.
This behavior is connected with the behavior of velocity of sound
with temperature through Eq.(\ref{pi1}). The initial decrease of $\zeta/s$
with temperature is due to increase of sound velocity at low temperature
due to excitation of light hadrons. At temperature $T>60$MeV, the sharp rise is related to the decrease of velocity of sound with excitations of heavier
hadrons leading to decrease of sound velocity which finally flattens out
at temperatures around 155 MeV as shown in Fig.\ref{thermo3}. The larger
bulk viscosity to entropy ratio at higher chemical potential is again related to
decrease of velocity of sound due to excitation of heavier baryons.

\begin{figure}[h]
\vspace{-0.4cm}
\begin{center}
\begin{tabular}{c c}
 \includegraphics[width=9cm,height=7cm]{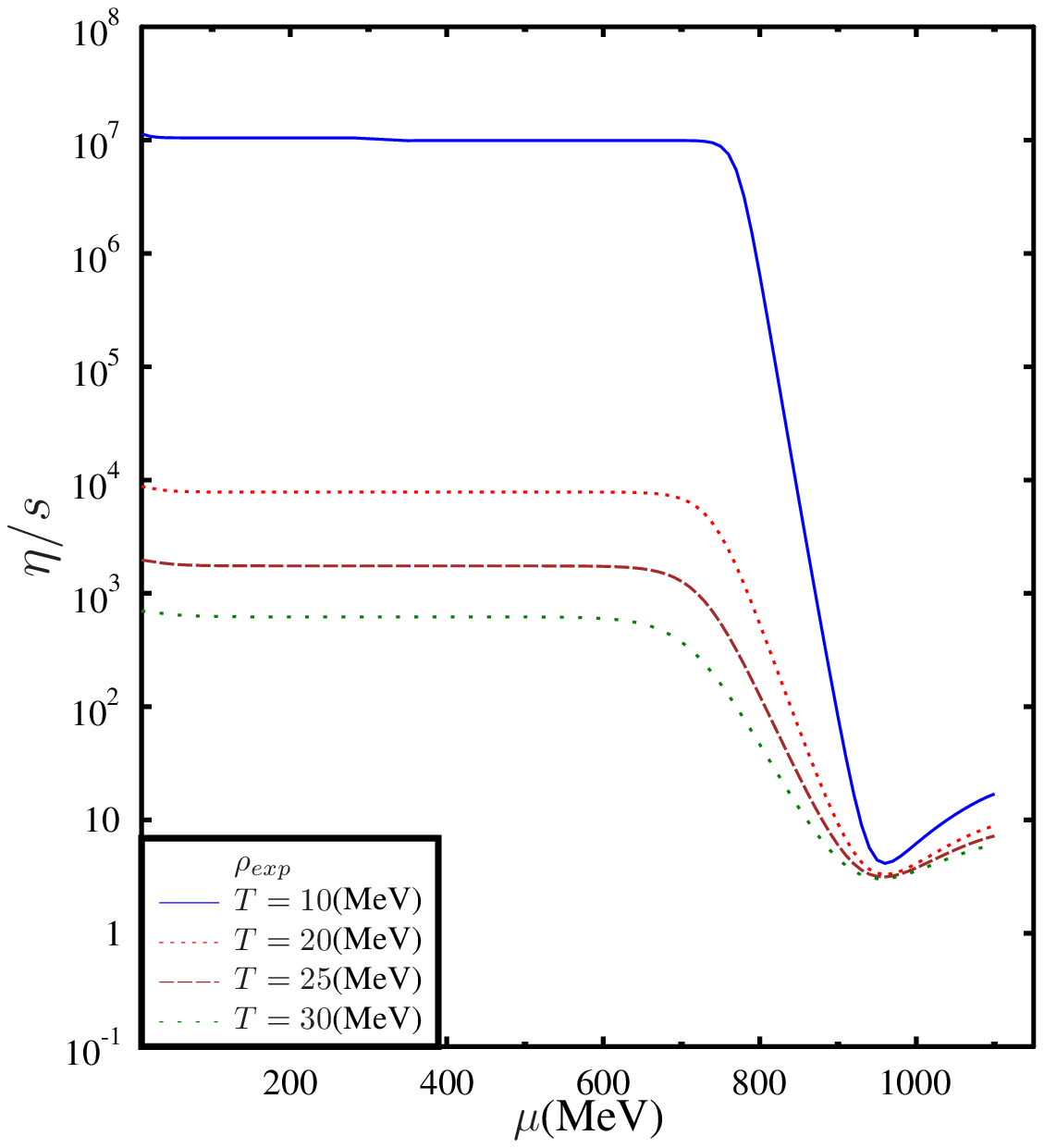}&
  \includegraphics[width=9cm,height=7cm]{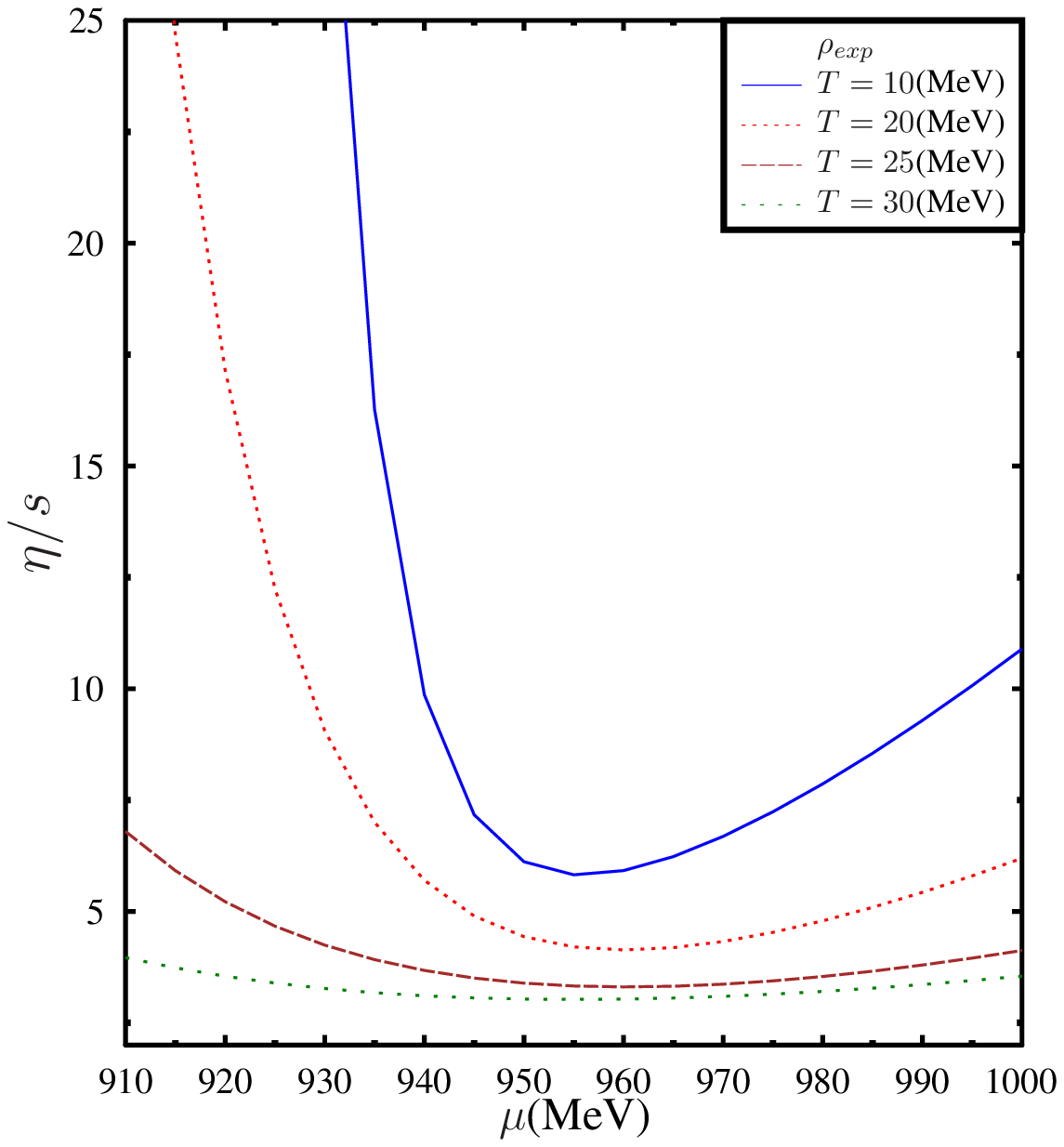}\\
 (7 a)&(7 b)
  \end{tabular}
\caption{Shear viscosity to entropy ratio  as a function of chemical potential.}
\label{etabismuu}
\end{center}
 \end{figure}

 In Fig.\ref{etabis}, we have plotted the shear viscosity to entropy ratio
for different chemical potentials as a function of temperature.
The finite volume effects arise here through the $\frac{1}{r^2}$ factors 
arising from the  finite size of the hadrons as in Eq.(\ref{eta}). 
We also retain here the finite volume corrections to the entropy
density $s$ as in Ref.\cite{peter}.  We might mention here that
the thermodynamic quantities are not sensitive to the
value of $r$ for $r> 0.2$fm as was demonstrated in Ref.\cite{greinerprc}. 
We have taken here a uniform size of $r=0.4 fm$ for all mesons
 and $r=0.5fm$ for all baryons \cite{bugaev,gorenstein}. 
For $\mu=0$ the minimum reaches about$\frac{\eta}{s} =0.7$ which is an order of magnitude larger than 
viscosity bound of $\frac{\eta}{s}=\frac{1}{4\pi}$. As the chemical potential is increased, temperature dependence is similar to that at $\mu=0$. On the other hand, the behavior of the ratio $\eta/s$ is nontrivial. It decreases
 with chemical potential for temperature less than about 130 MeV, beyond which
increases with $\mu$. The initial decrease of $\eta$ with respect to $\mu$
can be understood as enhancement of the hardcore cross section with nucleon
 number density . However, the entropy density $\s$ starts decreasing with 
increase in chemical potential at higher temperature.
 This is due to the fact that
the volume corrections proportional to the density of the particles
enter in the denominator for the entropy density \cite{peter}. This in turn
makes a larger value for the ratio $\frac{\eta}{s}$

We also looked into the behavior of $\frac{\eta}{s}$ at low temperatures
as a function of $\mu$ where it shows a valley structure which is plotted 
in Fig.\ref{etabismuu}. Such an observation was also made 
in Ref.s\cite{itakura, nakano,cpsingh}. The existence of a
minimum of $\eta/s$ was interpreted in these references indicative of a 
liquid gas phase transition. This is due to the fact that a minimum
in the ratio $\eta/s$  as a function of the controlling
parameter of thermodynamics like
temperature or chemical potential could be indicative of
a phase transition \cite{csernai,nakano,itakura}.
 As the temperature increase,
the valley structure become shallower as is clearly shown in Fig.7 b, 
possibly suggestive of the 
phase transition. However, on the other hand, the corresponding 
entropy does not show such a structure. Further, the corresponding nucleon
number density here (0.07/$fm^3$), however, turns out to be about
half the nuclear matter density.
\section{summary}
We have here tried to estimate the bulk and shear viscosity to entropy
ratio in a hadronic medium modeling the same as a hadron resonance gas. 
Apart from including all the hadrons  below a cutoff of 2 GeV, we have also included a continuum density of state beyond 2GeV. Such a description of hadronic model gives a good fit to the lattice data both at zero and finite chemical
potential \cite{borsonyimu}. The thermodynamic quantities so obtained is used to estimate the bulk viscosity of hadron gas at finite chemical potential using
the method as outlined in Ref.\cite{karschkharzeev} for finite temperature and zero chemical potential. At finite chemical potential,
the $\frac{\zeta}{s}$ become higher as compared to $\mu=0$ and is related to the fact that the velocity of sound becomes smaller due to finite chemical potential with excitation of heavier baryons contributing more to the energy density
as compared to the pressure.

This approach has already been used
to estimate $\eta/s$ for hadronic medium in Ref. \cite{greinerprl}
to obtain $\eta/s$ reaching the viscosity bound of $\frac{1}{4\pi}$ at 
temperature of about $T=190$MeV using the lattice data available at that time. 
However, later 
lattice data pointed to a lower critical temperature giving rise to indicate 
that the hadron resonance gas can lead to $\eta/s$ being about an order 
of magnitude higher than the viscosity bound. We  observe that 
at finite chemical potential $\frac{\eta}{s}$ increases with temperature 
with its magnitude increasing with chemical potential. For low temperatures $(T<30 MeV)$ and high baryon chemical potential, we observed a valley structure
for this ratio which can have a connection with liquid gas phase transition in nuclear matter.

\def\karschkharzeev{F. Karsch, D. Kharzeev, and K. Tuchin, Phys. Lett. B
663, 217 (2008).}
\def\joglekar{J.C. Collins, A. Duncan, S.D. Joglekar, Phys. Rev. D 16, 
438 (1977).}
\def\blaschke{J. Jankowski, D. Blaschke, M.Spalinski, Phys.Rev.D 87, 105018
(2013). }
\def\gorenstein{M. Gorenstein, M. Hauer, O. Moroz, Phys.Rev.C 77,024911 (2008)}
\def\bugaev{K. Bugaev et al, Eur.Phys.J. A 49, 30 (2013)}
\def\cpsingh{S.K. Tiwari, P.K. Srivastava, C.P. Singh, Phys.Rev. C 85,
014908 (2012) }
\def\chen{J.-W. Chen, Y-H. Li, Y.-F. Liu, and E. Nakano, Phys. Rev. D 76,
114011 (2007)}
\def\chennakano{J.-W. Chen, and E. Nakano, Phys. Lett. B 647, 371 (2007)}
\def\itakura{K. Itakura, O. Morimatsu, and H. Otomo, Phys. Rev. D 77, 014014
(2008)}
\def\cleymans{J. Cleymans, H. Oeschler, K. Redlich, and S. Wheaton, Phys.
Rev. C 73, 034905 (2006)}
\def\worku{J. Cleymans and D. Worku, Mod. Phys. Lett. A26,1197,(2011).}
\def\guptagod{S. Chatterjee, R. M. Godbole and S. Gupta, {\PRC{81}{044907}{2010}}.}
\def\Noronha{Noronha-Hostler J, Noronha J and Greiner C 2012 Phys. Rev. C 86 024913}
\def\berera{A. Bstero-Gil, A. Berera and R. Ramos, JCAP1107, 030 (2011).}
\def\heinzrev{U. Heinz and R. Snellings, Annu. Rev. Nucl. Part. Sci. 63, 123-151, 2013}
\def\hirano{P. Romatschke and U. Romatschke, Phys. Rev. Lett.{\bf 99},172301, (2007); T. Hirano and M. Gyulassy, Nucl. Phys. {\bf A 769}, 71, (2006).}
\def\kss{P. Kovtun, D.T. Son and A.O. Starinets, Phys. Rev. Lett.{\bf 94},
 111601, (2005).}
\def\tanmoy{A. Bazavov {\it etal}, e-print:arXiv:1407.6387.}
\def\cavitation{K. Rajagopal and N. Trupuraneni, JHEP1003, 018(2010);
 J. Bhatt, H. Mishra and V. Sreekanth, JHEP 1011, 106,(2010);{\it ibid} Phys. Lett. B704, 486 (2011); {\it ibid} Nucl. Phys. A875, 181(2012).}
\def\borsonyi{S. Borsonyi{\it etal}, JHEP1011, 077 (2010).}
\def\borsonyimu{S. Borsonyi{\it etal}, JHEP1208, 053 (2012).}
\def\dobado{A. Dobado,F.J.Llane-Estrada amd J. Torres Rincon, 
{\PLB{702}{43}{2011}}.}
\def\dobadoshear{A. Dobado,F.J.Llane-Estrada amd J. Torres Rincon, 
{\PRD{79}{055207}{2009}}.}
\def\sasakiqp{C. Sasaki and K.Redlich,{\PRC{79}{055207}{2009}}.}
\def\sasakinjl{C. Sasaki and K.Redlich,{\NPA{832}{62}{2010}}.}
\def\ellislet{I.A. Shushpanov, J. Kapusta and P.J. Ellis,{\PRC{59}{2931}{1999}}
; P.J. Ellis, J.I. Kapusta, H.-B. Tang,{\PLB{443}{63}{1998}}.}
\def\prakashwiranata{Anton Wiranata and Madappa Prakash,
{\PRC{85},{054908}{2012}}.}
\def\purnendu{P. Chakravarti and J.I. Kapusta {\PRC{83}{014906}{2011}}.}
\def\greco{S.Plumari,A. Paglisi,F. Scardina and V. Greco,{\PRC{83}{054902}{2012}a.}}
\def\bes{H. Caines, arXiv:0906.0305 [nucl-ex], 2009.}
\def\greinerprl{J. Noronha-Hostler,J. Noronha and C. Greiner, 
{\PRL{103}{172302}{2009}}.}
\def\greinerprc{J. Noronha-Hostler,J. Noronha and C. Greiner
, {\PRC{86}{024913}{2012}}.}
\def\igorgreiner{J. Noronha-Hostler, C. Greiner and I. Shovkovy,
, {\PRL{100}{252301}{2008}}.}
\def\majumdermueller{A. Majumder and B. Mueller, {\PRL{105}{252002}{2010}}.}
\def\leonidov{ A. V. Leonidov and P. V. Ruuskanen, {\EPJC{4}{519}{1998}}.}
\def\cbm{ B. Friman, C.H. Ohne, J. Knoll, S. Leupold, J. Randrup, R. Rapp, P. Senger (Eds.), Lect. Notes Phys., vol. 814,
2011.}
\def\nica {A.N. Sissakian, A.S. Sorin, J. Phys. G 36 (2009) 064069.}
\def\nakano{J.W. Chen,Y.H. Li, Y.F. Liu and E. Nakano,
 {\PRD{76}{114011}{2007}}.}
\def\itakura{K. Itakura, O. Morimatsu, H. Otomo, {\PRD{77}{014014}{2008}}.}
\def\wang{M.Wang,Y. Jiang, B. Wang, W. Sun and H. Zong, Mod. Phys. lett.
{\bf A76}, 1797,(2011).}
\def\agasian{N.O. Agasian, JETP Lett. 95, 171, (2012), arXiv:1109.5849.}
\def\Hagedorn{R. Hagedorn and J. Rafelski,{\PLB{97}{136}{1980}}.}
\def\kapustaolive{J.I. Kapusta and K. A. Olive, {\NPA{408}{478}{1983}}.}
\def\hrgexp{P. Braunmunzinger, J. Stachel, J.P. Wessels and N. Xu,
{\PLB{365}{1}{1996}}; G.D. Yen and M.I. Gorenstein, {\PRC{59}{2788}{1999}};
F. Becattini, J. Cleymans, A. Keranen, E. suhonen and K. Redlich, 
{\PRC{64}{024901}{2001}}.}
\def\rischkegorenstein{.D.H. Rischke, M.I. Gorenstein, H. Stoecker and
W. Greiner, Z.Phys. C {\bf 51}, 485 (1991).}
\def\hmnjl{Amruta Mishra and Hiranmaya Mishra, {\PRD{74}{054024}{2006}}.}
\def\pdgb{C. Amseler {\it et al}, {\PLB{667}{1}{2008}}.}
\def\shuryak{E.V. Shuryak, Yad. Fiz. {\bf 16},395, (1972).}
\def\leupold{S. Leupold, J. Phys. G{\bf32},2199,(2006)}
\def\peter{A. Andronic, P. Braun-Munzinger , J. Stachel and M. Winn,
{\PLB{718}{80}{2012}}}
\def\blum{M. Blum, B. Kamfer, R. Schluze, D. Seipt and U. Heinz,{\PRC{76}{034901}{2007}}.}
\def\jaminplb{M. Jamin{\PLB{538}{71}{2002}}.}
\def\csernai{L.P. Csernai, J.I. Kapusta and L.D. McLerran,{\PRL{97}{152303}{2006}}.}
\def\hagedorn{R. Hagedorn, Nuovo Cim. Suppl. 3,147 (1965); Nuovo Sim. A56,1027 (1968).}


\begin{thebibliography}{99}
\bibitem{heinzrev}\heinzrev
\bibitem{berera}\berera
\bibitem{hirano}\hirano
\bibitem{kss}\kss
\bibitem{tanmoy}\tanmoy
\bibitem{borsonyi}\borsonyi
\bibitem{cavitation}\cavitation
\bibitem{dobado}\dobado
\bibitem{sasakiqp}\sasakiqp
\bibitem{sasakinjl}\sasakinjl
\bibitem{karschkharzeev}\karschkharzeev
\bibitem{ellislet}\ellislet
\bibitem{dobadoshear}\dobadoshear
\bibitem{prakashwiranata}\prakashwiranata
\bibitem{purnendu}\purnendu
\bibitem{greco}\greco
\bibitem{gorenstein}\gorenstein
\bibitem{greinerprl}\greinerprl
\bibitem{hrgexp}\hrgexp
\bibitem{greinerprc}\greinerprc
\bibitem{bes}\bes
\bibitem{cbm}\cbm
\bibitem{nica}\nica
\bibitem{nakano}\nakano
\bibitem{itakura}\itakura
\bibitem{wang}\wang
\bibitem{agasian}\agasian
\bibitem{hagedorn}\hagedorn
\bibitem{kapustaolive}\kapustaolive
\bibitem{rischkegorenstein}\rischkegorenstein
\bibitem{majumdermueller}\majumdermueller
\bibitem{leonidov}\leonidov
\bibitem{igorgreiner}\igorgreiner
\bibitem{guptagod}\guptagod
\bibitem{cleymans}\cleymans
\bibitem{Hagedorn}\Hagedorn
\bibitem{worku}\worku
\bibitem{shuryak}\shuryak
\bibitem{borsonyimu}\borsonyimu
\bibitem{blaschke}\blaschke
\bibitem{jaminplb}\jaminplb
\bibitem{leupold}\leupold
\bibitem{hmnjl}\hmnjl
\bibitem{pdgb}\pdgb
\bibitem{blum}\blum
\bibitem{peter}\peter
\bibitem{bugaev}\bugaev
\bibitem{csernai}\csernai

\bibitem{cpsingh}\cpsingh
\vfil
\end{thebibliography}
\end{document}